\documentclass[published]{JHEP3} 

\JHEP{00(2004)000}

\JHEPspecialurl{http://jhep.sissa.it/JOURNAL/JHEP3.tar.gz}

\usepackage{epsfig,multicol,amsmath,amssymb,amsfonts}

\newcommand\fverb{\setbox\pippobox=\hbox\bgroup\verb}
\newcommand\fverbdo{\egroup\medskip\noindent%
            \fbox{\unhbox\pippobox}\ }
\newcommand\fverbit{\egroup\item[\fbox{\unhbox\pippobox}]}
\newbox\pippobox

\title{On infinite-dimensional representations of the rotation group and Dirac monopole}

\author{Alexander I. Nesterov and F. Aceves de la Cruz\\
  E-mail: \email{nesterov@cencar.udg.mx}, \email{fermin@udgphys.intranets.com}\\

    Departamento de F{\'\i}sica, CUCEI, Universidad de
Guadalajara, Av. Revoluci\'on 1500, Guadalajara, CP 44420, Jalisco,
M\'exico\\
}
\received{}        
\accepted{}        

\preprint{\hepth{9912999}}  

\abstract{The Dirac monopole problem is studied in details within the framework of infinite-dimensional
representations of the rotation group, and a consistent pointlike monopole theory with an arbitrary
magnetic charge is deduced.
}


\begin{document}

\section{Introduction}

In 1931 Dirac \cite{Dir} showed that a proper description of the quantum
mechanics of a charged particle of the charge $e$ in the field of the
magnetic monopole of the charge $q$ requires the quantization condition
$2\mu =n, \, n \in \mathbb Z$ (we set $eq =\mu$ and $\hbar = c =1$). Well-known group theoretical, topological and geometrical arguments in behalf of Dirac  quantization rule are based on employing classical fibre bundle theory  or finite dimensional representations of the rotation group \cite{Dir,Sw_1,Wu1,Wu2,Jac,Gr,Gr1,G1,G2,G3}.

For instance, a realization of the Dirac monopole as U(1) bundle over
$S^2$ implies that there exists the division of space into overlapping
regions $\{U_i\}$ with nonsingular vector potential being defined in
$\{U_i\}$, and yields the correct monopole magnetic field in each region.
On the triple overlap $U_i\cap U_j\cap U_k$ it holds
\begin{equation}
\exp(i(q_{ij}+q_{jk}+q_{ki})) =\exp(i4\pi\mu)
\end{equation}
where $q_{ij}$ are the transition functions, and the consistency
condition, which is equivalent to the associativity of the group
multiplication, requires $q_{ij}+q_{jk}+q_{ki} = 0 \mod2\pi \mathbb Z$.
This yields the Dirac selectional rule $2\mu \in \mathbb Z$ as a necessary
condition to have a consistent U(1)-bundle over $S^2$ \cite{Wu1,Wu2,Gr}.

In the presence of the magnetic monopole the operator of the total angular momentum ${\mathbf J}$, which uncludes contribution of the electromagnetic field,  obeys the standard commutation relations of the Lie algebra of the rotation group
\begin{eqnarray*}
[J_i, J_j] = i\epsilon_{ijk}J_k \label{eq4a},
\end{eqnarray*}
and this is true for any value of $\mu$. However, the requirement that $J_i$ generate a finite-dimensional representation of the rotation group yields $2\mu$ being integer and only values $2\mu = 0, \pm1,\pm2, \dots$ are allowed \cite{Ch,Gol1,Gol2,Zw_1,Zw_2,H}.

Thus to avoid the Dirac restrictions on the magnetic charge one needs to consider a nonassociative generalization of U(1) bundle over $S^2$ and give up finite-dimensional representations of the rotation group. Recently we have done the first steps in this direction, developing a consistent pointlike monopole theory with an arbitrary magnetic charge \cite{NF,N1,N2}. Here we study in details the Dirac monopole problem within the framework of infinite-dimensional representations of the rotation group.

The paper is organized as follows. In Section II  the indefinite metric
Hilbert space is introduced. In Section III the properties of
infinite-dimensional representations of the rotation group are discussed.
In Section IV it is argued that expanding the representations of the
rotation group to infinite-dimensional representations allow an arbitrary
magnetic charge. In Section V the obtained results and open problems are
discussed.

\section{Indefinite metric Hilbert space.}

Starting from the early 1940s indefinite metric in the Hilbert space has been
discussed and used by many authors. Recently a growning interest to this topic
has been rised in the context of the so-called PT-symmetric quantum
mechanics related with some non-Hermitian Hamiltonians with a real spectrum and
pseudo-Hermitian operators \cite{BB,BBM,BBJ,M1,M2,M3,M4,M5,Sol,BSS,RM}.

Since in conventional quantum mechanics the norm of quantum states given by
\begin{equation}\label{n}
 \int \bar\psi \psi d x > 0
\end{equation}
where $\bar \psi$ is the conjugate complex of $\psi$, carries a probabilistic interpretation, the appearance of an indefinite metric in Hilbert space is a sever obstacle. It leads in particular to {\em negative probability of states}, that means observables with only positive eigenvalues can get negative expectation values \cite{Pauli}.

We treat here the more general situation when the normalization given by
\begin{equation}\label{n1}
 \int \bar\psi \psi d \mu(x),
\end{equation}
$d \mu$ being a suitable measure, is not necessary positive. We assume that the integral
\begin{equation}\label{n1a}
 \int \bar\psi \psi' d \mu(x),
\end{equation}
may be divergent and its value is given by a regularization (for the definition of regularization of an integral see, {\em e.g.} \cite{GSH})\footnote {There exist several possibilities of regularizing divergent integral, further (see Sect. 3) we consider the regularization of the integral by analytical continuation in parameter.}.

Following the notations by Pauli \cite{Pauli} we consider an {\em inner product} in the indefinite metric
Hilbert space $\mathcal H^\eta$ defined by the bilinear form of the type \begin{equation}
(\bar\psi,\psi')_\eta = \int \bar\psi {\eta} \psi' d \mu(x),
\label{h1}
\end{equation}
in which the operator $\eta$ is only restricted by the condition that it has to be Hermitian and
\begin{equation}\label{n2}
 \int \bar\psi {\eta} \psi d \mu(x) > 0.
\end{equation}
The difference between our construction of the indefinite Hilbert space and suggested in \cite{Pauli}
arises from the restrictions (\ref{n1}) and (\ref{n2}). While Pauli requires the positive defined norm
(\ref{n1}), we don't.

Let functions $\psi_m(x)$ form the basis such that
\begin{eqnarray*}
\int \overline \psi_m(x)\psi_{m'}(x)d\mu(x)= \eta_{mm'}
\end{eqnarray*}
where  $\eta_{m m'}= (-1)^{\sigma(m)}\delta_{mm'}$ is an indefinite diagonal metric $(-1)^{\sigma(m)} = \pm
1$ depending whether $\sigma(m)$ is even or odd. Defining the action of the operator $\eta $ on $\psi_m$
as
\begin{equation}\label{eta1}
    \eta\psi_m = \eta_{mm'}\psi_{m'}
\end{equation}
we find that the set $\{\psi_m\}$ forms the orthonormal basis with respect to the inner product given by
\begin{equation}\label{eta2}
(\psi_m,\psi_{p})_\eta = (\psi_m,\eta\psi_{p}) = \int \overline \psi_m(x)\eta_{pm'}\psi_{m'}(x)d\mu(x)
=\delta_{mp}.
\end{equation}

Since the set $\{\psi_m(x)\}$ forms a basis, an arbitrary function
$\psi(x)\in\mathcal H$ can be expanded in terms of the $\psi_m(x)$:
\begin{eqnarray}
\psi(x) = \sum_m c^\eta_m \psi_m,
\end{eqnarray}
where
\begin{equation}\label{N2}
   c^\eta_m = (\psi_m, \psi)_\eta = \eta_{mm'}\int \overline \psi_{m'}(x)\psi(x)d\mu(x)
\end{equation}

Let $ \psi'(x)\in \mathcal H$, which can be expanded as follows:
\begin{align}
    &\psi'(x) = \sum_m c'^\eta_m \psi_m,
   \end{align}
then the inner product $(\psi,\psi')_\eta$ can be easy calculated that is
\begin{equation}
(\psi,\psi')_\eta = \int \bar \psi(x){\eta}\psi'(x)d\mu(x) = \sum_m
\overline c^\eta_m c'^\eta_m.
\label{h3}
\end{equation}
In particular:
\begin{equation*}
 (\psi,\psi)_\eta = \int \bar \psi(x){\eta}\psi(x)d\mu(x) = \sum_m |c^\eta_m|^2
>0.
\end{equation*}
Thus  we see that the inner product in the indefinite metric Hilbert space is
positive defined scalar product. This provides the standard probabilistic
interpretation of the quantum mechanics.

The inner product ({\ref{h3}}) may be written in another form. Let us consider the sum
\begin{equation}\label{k_3}
K(x,x') = \sum_m \psi_m(x)\overline {\psi_m(x')}.
\end{equation}
This yields the following relations:
\begin{align}
 & \int \psi_{m}(x')K(x,x') d\mu(x')= \eta_{m m'}\psi_{m'}(x),
  \label{k_2}\\
&{\eta}\psi'(x)=  \int \psi'(x')K(x,x')d\mu(x'),
\label{k_2a}
\end{align}
and it is seen that the kernel $K(x,x')$ plays here a role similar to that of
$\delta$-function in the standard Hilbert space of quantum mechanics. Now one
can rewrite the inner product (\ref{h3}) as
\begin{equation}
(\psi,\psi')_\eta = \int \int\bar \psi(x)K(x,x')\psi'(x')d\mu(x)d\mu(x').
\end{equation}

The expectation value of the an observable $A$ represented by the linear operator acting in $\mathcal H$ is
defined by
\begin{equation}\label{A1}
    \langle A\rangle_\eta =\int \bar \psi(x){\eta}A\psi(x)d\mu(x),
\end{equation}
and the generalization of the Hermitian conjugate operator, being denoted as $A^\dag_\eta$, is given by
\begin{equation}\label{A2}
A^\dag_\eta = \eta^{-1}A^\dag \eta
\end{equation}
where $A^\dag$ is the Hermitian conjugate operator.

Since the observables are real, we see that for them the related operators have
to be self-adjoint in the indefinite metric Hilbert space, that means
$A^\dag_\eta= A$. In particular, applying this to the Hamiltonian operator $H$,
we have $ H^\dag_\eta= H$, and assuming that the wave function satisfies the
Schr\"odinger's equation
\[
i\frac{\partial{\psi}}{\partial{t}} = H\psi
\]
we obtain
\begin{equation}\label{A3}
    \frac{d}{dt}(\psi|\psi)_\eta = i\bar\psi \eta(H^\dag_\eta- H)\psi =0,
\end{equation}
that is, the conservation of the wave function normalization.

If we perform a linear transformation  $$\psi' = S\psi$$
then in order to conserve the normalization of the wave function
\[
(\psi',\psi')_\eta = (\psi,\psi)_\eta
\]
we have to demand
\begin{equation}\label{A4}
    \eta' = S^\dag\eta S.
\end{equation}
In similar manner we find that the observables are invariant,
\begin{equation}\label{A5}
\langle A'\rangle_\eta =\langle A\rangle_\eta,
\end{equation}
if the operators transform as follows:
\begin{equation}\label{A5a}
    A'= S^{-1}A S, \quad A^{\dag'}_\eta= S^{-1}A^\dag_\eta S.
\end{equation}

Assuming that, according to (\ref{A5a}), the matrix $A$ can be transformed with
a suitable $S$ to a normal form  such that
\begin{equation}\label{A6}
A\psi_n = a_n\psi_n
\end{equation}
we  find
\begin{equation}\label{A6a}
(\psi,A\psi)_\eta = \sum_n a_n|c^\eta_n|^2.
\end{equation}
This leads to the conclusion that {\em operator with only positive eigenvalues
can not have negative expectation values }. In other words, the negative
probabilities do not appear in our approach, and the standard interpretation of
the wave function normalization is preserved.

\section{Infinite-dimensional representations of the rotation group}

The three dimensional rotation group is locally isomorphic  to the
group SU(2), and as well known SO(3)=${\rm SU(2)}/\mathbb{Z}_2$. In
what follows the difference between SO(3) and SU(2) is not
essential and actually we will consider  G=SU(2). The Lie algebra
corresponding to the Lie group SU(2) has three generators and we
adopt the basis $J_{\pm} = J_{1} \pm iJ_{2}, \; J_3$.  The
commutation relations are:
\begin{equation}
[J_{+}, J_{-}] = 2J_3, \quad [J_3, J_{\pm}] = \pm J_{\pm}\quad
[J^2 , J_{\pm}]=0 ,\quad [J^2,J_3] =0
 \label{J_1}
\end{equation}
where
\begin{eqnarray}
J^2=J_3^2 +\frac{1}{2}(J_{-}J_{+}+J_{+}J_{-}) \label{Cas}
\end{eqnarray}
is the Casimir operator.

Let $\psi^\lambda_\nu$ be an eigenvector of the operators $J_3$ and $J^2$:
\begin{equation}\label{J1a}
J_3\psi^\lambda_\nu=(\nu_0 +n)\psi^\lambda_\nu, \quad J^2\psi^\lambda_\nu =\lambda(\lambda
+1)\psi^\lambda_\nu,
\end{equation}
where $n=0,\pm 1,\pm 2,\dots$, and $\nu_0$, just like $\lambda$, is a certain complex number.

There are four distinct classes of representations and each irreducible
representation is characterized by an eigenvalue of Casimir operator and the
spectrum of the operator $J_3$ \cite{AG,BL,S,S1,S2,W}:
\begin{itemize}
\item {\em Representations unbounded from above and below}, in this case neither $\lambda + \nu_0$ nor $\lambda - \nu_0$ can be integers.

\item {\em Representations bounded below}, with $\lambda + \nu_0$ being an integer, and $\lambda - \nu_0$ not equal to an integer.

\item {\em Representations bounded above,} with $\lambda - \nu_0$ being an integer, and $\lambda + \nu_0$ not equal to an integer.

\item {\em Representations bounded from above and below,} with $\lambda - \nu_0$ and $\lambda + \nu_{0}$ both being integers, that yields $\lambda = k/2, \quad k\in \mathbb Z_{+}$.

\end{itemize}
The nonequivalent representations in the each series of irreducible
representations are denoted respectively by $D(\lambda,\nu_0)$,
$D^{+}(\lambda)$, $D^{-}(\lambda)$ and $D(\lambda)$. The representations
$D(\lambda,\nu_0)$, $D^{+}(\lambda)$ and $D^{-}(\lambda)$ are infinite-
dimensional; $D(\lambda)$ is $(2\lambda+1)$-dimensional representation. The irreducible
representations $D^{\pm}(\lambda)$ and $D(\lambda,\nu_0)$ are discussed in
details in \cite{AG,BL,S,S1,S2}. Further we restrict ourselves by the real
eigenvalues of the Casimir operator and $J_3$ denoting their by $\ell$ and $m$:
$\lambda \rightarrow\ell$ and $\nu \rightarrow m$.

\subsection*{Representations unbounded from above and below}

Let $\psi_m$ be non normalized eigenstates  of the operators $J_3$ and $J^2$:
\[
J_3\psi_m = m \psi_m, \quad J^2\psi_m = \ell(\ell + 1) \psi_m.
\]
Demand that the commutation relations (\ref{J_1}) satisfied, yields
\begin{eqnarray}
&&J _{-} \psi_m = (\ell +m)\psi_{m-1}, \label{J_1b} \\
&&J _{+}\psi_m  = (\ell - m)\psi_{m+1}, \label{J_2a}. \label{J_3b}
\end{eqnarray}

Considering the invariance of an inner product $(\psi_m,\psi_{m'})$ with
respect to infinitesimal rotations generated by $J_i$ we obtain
\begin{eqnarray}\label{rot_1}
&&(\psi_m,(J_+ - J_-)\psi_{m'}) +(\psi_m(J_+ - J_-),\psi_{m'}) =0,\\
&&(\psi_m,\psi_{m'}) =0, \quad m \neq m'.
\end{eqnarray}
Putting $m' = m+1$ we find
\begin{equation}\label{rot_2}
(\psi_m,J_-\psi_{m+1}) -(\psi_m J_+,\psi_{m+1}) =0,
\end{equation}
that yields the following restriction on the inner product:
\begin{equation}\label{rot_2a}
(l+m +1)(\psi_m,\psi_{m}) - (l -m )(\psi_{m+1},\psi_{m+1}) =0.
\end{equation}
This recursion relationship can be satisfied by writing
\begin{equation}\label{pr_1}
(\psi_m,\psi_{m})=\mathcal {N}\,\Gamma(\ell +m +1)\Gamma(\ell - m + 1)
\end{equation}
where $\mathcal {N}$ is an arbitrary positive constant, $\Gamma$ is the gamma
function, and for $\ell \pm m +1 < 0$ the value of r.h.s. is given by
analytical continuation of the gamma function.

Introducing
\begin{equation}\label{N_1} {\mathcal N}_m = (\mathcal
{N}\,|\Gamma(\ell +m +1)\Gamma(\ell - m + 1)|)^{-\frac{1}{2}},
\end{equation}
we obtain
\begin{equation}\label{pr_1a} \mathcal
{N}^2_m(\psi_m,\psi_{m})=(-1)^{\sigma(m)},
\end{equation}
where
\begin{eqnarray*}
 (-1)^{\sigma(m)}= {\rm sgn}\big(\Gamma(\ell-m +1)\Gamma(\ell+m+1)\big),
\end{eqnarray*}
${\rm sgn}(x)$ being the signum function.

It follows from Eq. (\ref{pr_1a}) that the states ${|\ell,m
\rangle}= {\mathcal N}_m\psi_m $ form the orthonormal basis under
the inner product given by
\begin{equation}\label{nor_1}
\langle m,\ell|\ell,m'\rangle_{\eta} = {\mathcal N}^2_m \eta_{m
m'}(\psi_m,\psi_{m'})= \delta_{m m'},
\end{equation}
with the indefinite metric  being $\eta_{m m'} = (-1)^{\sigma(m)}\delta_{m
m'}$.

We found that the operators $J_{\pm}$ act on the states $|\ell,n\rangle$ as follows:\\

$\bullet $ for $-\ell < m < \ell $
\begin{align}
& J_{+}|\ell,m\rangle = \sqrt{\ell(\ell +1) - m(m +
1)} \;|\ell ,m + 1\rangle  \label{rep_1a}\\
&J_{-}|\ell,m\rangle =  \sqrt{\ell(\ell +1) - m(m - 1)} \;|\ell ,m
-1\rangle
\end{align}

 $\bullet $ for $|m| > \ell$
\begin{align}
&J_{+}|\ell,m\rangle = \sqrt{m(m+1) - \ell(\ell +1)
} \;|\ell ,m + 1\rangle  \\
&J_{-}|\ell,m\rangle = - \sqrt{m(m -1) - \ell(\ell +1)} \;|\ell ,m
-1\rangle \label{rep_1b}
\end{align}
and for the matrix elements one has
\begin{eqnarray*}
{(J_{+})}_{m'm}= \left\{\begin{array}{l}
\overline{(J_{-})}_{m m'},\;\textrm {if}\; -\ell < m < \ell  \\
-\overline{(J_{-})}_{m m'},\;\textrm {if}\; |m| > \ell \\
\end{array}\right.
\end{eqnarray*}
where `bar' denotes complex conjugation.

One can start with an arbitrary vector $|\ell,m\rangle$ and $\mu$ being an arbitrary
number with the fixed value within the given irreducible representation, and apply the
operators $J_{\pm}$ to  obtain any state $|\ell,m'\rangle$. Since the eigenvalues of $J_3$ can be changed only by multiples of unity, one has $m = \mu +p$, $p \in \mathbb Z$.
Thus each irreducible representation $D({\ell},\mu) $ may be characterized by the given values of two invariants $\ell$ and $\mu$. In fact the representations $D{(\ell},\mu) $ and $D({-\ell -1},\mu) $, yielding the same value  $Q= \ell(\ell +1)$ of the Casimir operator, are equivalent and the inequivalent representations may be labeled as $D({Q},\mu) $ \cite{W}.

If there exists the number $p_0$ such that $\mu + p_0= \ell$, we have $J_{+}|\ell,\ell\rangle = 0$ and the representation becomes bounded above. In the similar manner if for a number $p_1$ one has $\mu + p_1=-\ell $, then $J_{-}|\ell,-\ell\rangle = 0$ and the representation reduces to the
bounded below. Finally, finite-dimensional unitary representation arises when there exist possibility of finding $J_{+}|\ell,\ell\rangle = 0$ and $J_{-}|\ell,-\ell\rangle = 0$. It is easy to see that in this case $2\ell, \;2m$ and $2 \mu$ all must be integers.

\subsection*{Representations bounded above}

It is convenient to set $m =\ell -n$ and consider the orthonormal states
$|\ell,n\rangle$  instead of $|\ell,m\rangle$. The vectors
$|\ell,n\rangle$ form a basis in the space of the representation
$D^+({\ell})$, where the operator $J_3$ acts as follows:
\begin{equation}
J_3|\ell,n\rangle=(\ell-n)|\ell,n\rangle, \quad n=0,1,\dots,\infty.
\end{equation}
This representation is characterized by the eigenvalue $\ell$ of the
highest-weight state:  $|\ell,0\rangle$ such that $J_{+}|\ell,0\rangle =
0$ and $ J_{3} |\ell,0\rangle = \ell|\ell,0\rangle$. The action of the
operators $\{J_{\pm}\}$ on the states is given by
\begin{align*}
&\left.\begin{array}{l} J _{+} |\ell,n\rangle =
\sqrt{n(2\ell-n+1))}|\ell,n-1\rangle \\
J _{-} |\ell,n\rangle = \sqrt{(n+1) (2\ell -n)}|\ell,n+1\rangle
  \\
\end{array}\right \} \; 0\leq n< 2\ell,\\
&\left.\begin{array}{l} J _{+} |\ell,n\rangle = \sqrt{n(n -2\ell
-1))}|\ell,n-1\rangle
 \\
J _{-} |\ell,n\rangle = -\sqrt{(n+1) (n-2\ell)}|\ell,n+1\rangle
  \\
\end{array}\right \}\; n> 2\ell
\end{align*}

We consider a suitable realization of the representation $D^{+}({\ell})$ in
the space of entire analytical functions ${\mathcal F}^{\ell} = \{f(z): z
\in \mathbb C\}$. In this realization the generators $J_{\pm}$ and $J_3$
act as the first order differential operators:
\begin{align}
J_{-}= -z^2\partial_{z} + 2\ell z,\; J_{+}=\partial_{z},\; J_{3}=
-z\partial_{z} + \ell,
\end{align}
The monomials
\[
\langle z|\ell,n\rangle = {\mathcal N}_n z^n,
\]
where ${\mathcal N}_n = {(\Gamma(n+1)|\Gamma(2\ell - n + 1)|)}^{-1/2}$ is
the normalization constant, form an orthogonal basis for holomorphic
functions analytical in $\mathbb C$, and satisfy
\begin{eqnarray}
(z^n,z^p): =\frac{\Gamma(2\ell + 2)}{2\pi i}\int \frac{
\bar{z}^n z^p d\bar z d z }{(1+|z|^2)^{2\ell +2}}= \Gamma(n+1)\Gamma(2\ell - n + 1)\delta_{np}.
\label{p1b}
\end{eqnarray}
For $n > 2\ell$ the value of r.h.s. is given by the analytical continuation of
the gamma function \cite{S}.

It follows from Eq.(\ref{p1b}) that the states $|\ell,n\rangle$ form the
orhonormal basis under the indefinite metric inner product defined as
follows:
\begin{eqnarray}
\langle n,\ell |\ell,p\rangle_\eta = \eta_{pp'}(\langle z|\ell,n\rangle,\langle
z|\ell,p'\rangle) = \delta_{np},
\end{eqnarray}
where $\eta_{np}= (-1)^{\sigma(n)}\delta_{np}$ and
\begin{align*}
&(-1)^{\sigma(n)}=\left\{\begin{array}{l} 1,\; {\rm if}\; 2\ell -n >0\\
(-1)^{n+1}\,{\rm sgn}(\sin2\pi\ell),\; {\rm if}\; n -2\ell >0
  \end{array}\right.
\end{align*}

An arbitrary state of the representation is an entire function of the type
\begin{equation}\label{f_1}
f(z) = \sum^{\infty} _{n=0}f_n \langle z|\ell,n\rangle.
\end{equation}
The inner product of two entire functions $f(z)$ and $g(z)$ is constructed
as follows:
\begin{eqnarray}\label{pr_2}
\langle f|g\rangle_\eta =  \frac{\Gamma(2\ell +2)}{2\pi i}\int_D
\frac{ \bar{f} {\eta }g\,d\bar z d z}{(1+|z|^2)^{2+2\ell}},
\end{eqnarray}
where the action of the operator $\eta$ is given by
\begin{equation}\label{k_3ab}
{\eta}|\ell,n\rangle) = \eta_{np}|\ell,p\rangle).
\end{equation}

\subsection*{Representations bounded below}

For the representation bounded below, setting $m=n - \ell$, we have
\begin{equation}
J_3|\ell,n\rangle=(n -\ell)|\ell,n\rangle, \quad n=0,1,\dots,\infty.
\end{equation}
The action of the operators $\{J_{\pm}\}$ on the states is given by
\begin{align*}
&\left.\begin{array}{l} J _{-} |\ell,n\rangle =
\sqrt{n(2\ell-n+1))}|\ell,n-1\rangle \\
J _{+} |\ell,n\rangle = \sqrt{(n+1) (2\ell -n)}|\ell,n+1\rangle
  \\
\end{array}\right \} \; 0\leq n< 2\ell,\\
&\left.\begin{array}{l} J _{-} |\ell,n\rangle = \sqrt{n(n -2\ell
-1))}|\ell,n-1\rangle
 \\
J _{+} |\ell,n\rangle = -\sqrt{(n+1) (n-2\ell)}|\ell,n+1\rangle
  \\
\end{array}\right \}\; n> 2\ell
\end{align*}

The representation is characterized by the eigenvalue $\ell$ of the
highest-weight state:  $|\ell,0\rangle$ such that $J_{-}|\ell,0\rangle =
0$ and $ J_{3} |\ell,0\rangle = -\ell|\ell,0\rangle$.

We consider a  realization of the representation $D^{-}({\ell})$ in the
space of analytical functions ${\mathcal F}^{\ell} = \{f(z): z\in \mathbb
C\}$, such that $z^{-2\ell}f(z)$ is the meromorphic function. In this
realization the generators $J_{\pm}$ and $J_3$ act as the following
differential operators:
\begin{align}
J_{-}= -z^2\partial_{z} + 2\ell z,\; J_{+}=\partial_{z},\; J_{3}=
-z\partial_{z} + \ell,
\end{align}
The monomials
\begin{equation}
\langle z|\ell,n\rangle = {\mathcal N}_n z^{2\ell- n}, \\
\end{equation}
${\mathcal N}_n = {(n!|\Gamma(2\ell - n + 1)|)}^{-1/2}$ being the same normalization constant as above,
form an orthonormal basis such that
\begin{eqnarray*}
\langle n,\ell |\ell,p\rangle_\eta
=\frac{(-1)^{\sigma(n)}\Gamma(2\ell + 2)}{n!|\Gamma(2\ell - n +
1)|}\frac{1}{2\pi i}\int  \frac{ \bar{z}^{2\ell-n} z^{2\ell-p}d\bar z d
z}{(1+|z|^2)^{2\ell+2}}= \delta_{np}
\end{eqnarray*}
where
\begin{align*}
&(-1)^{\sigma(n)}=\left\{\begin{array}{l} 1,\; {\rm if}\; 2\ell -n >0\\
(-1)^{n+1}\,{\rm sgn}(\sin2\pi\ell),\; {\rm if}\; n -2\ell >0
  \end{array}\right.
\end{align*}

An arbitrary state of the representation is a function of the type
\begin{equation}\label{f_1a}
f(z) = \sum^{\infty} _{n=0}f_n \langle z|\ell,n\rangle.
\end{equation}
The inner product of the functions $f(z)$ and $g(z)$ is constructed as above (see Eq.(\ref{pr_2})):
\begin{eqnarray}\label{p2e}
\langle f|g\rangle_\eta =  \frac{\Gamma(2\ell +2)}{2\pi i}\int \frac{ \bar{f}
{\eta }g d\bar z d z}{(1+|z|^2 )^{2+2\ell}}.
\end{eqnarray}

\section{Infinite-dimensional representations and Dirac monopole problem}

For a non relativistic charged particle in the field of a magnetic
monopole the equations of motion
\begin{equation}
m\ddot{\mathbf r} = -\frac{\mu}{r^3}{\mathbf r} \times\dot{\mathbf r}
\label{eq1}
\end{equation}
imply that the total angular momentum
\begin{equation}
{\mathbf J} = {\mathbf r} \times \left({\mathbf p} - e{\mathbf A}\right) -
\mu\frac{\mathbf r}{r} \label{eq1c}
\end{equation}
is conserved. The operator of the angular momentum
\begin{equation}
{\mathbf J} = {\mathbf r} \times \left({-i\mathbf \nabla} - e{\mathbf
A}\right) - \mu\frac{\mathbf r}{r},
\end{equation}
having the same properties as a standard angular momentum, obeys the following
commutation relations:
\begin{eqnarray}
&&[H, {\mathbf J}^2] = 0, \quad [H, J_i] = 0,\quad  [{\mathbf J}^2, J_i] =
0, \label{eq5a} \\
&&[J_i, J_j] = i\epsilon_{ijk}J_k \label{eq5}
\end{eqnarray}
where $H$ is the Hamiltonian.

As well known any choice of the vector potential $\mathbf A$ being
compatible with a magnetic field ${\mathbf B}$ of Dirac monopole
must have  singularities (the so-called strings), and one can write
\[
{\mathbf B}={\rm rot}{\mathbf A}_{\mathbf n} + {\mathbf h}_{\mathbf n}
\]
where ${\mathbf h}_{\mathbf n}$ is the magnetic field of the string.

For instance, Dirac introduced the vector potential as
\begin{equation}
{\mathbf A}_{\mathbf n}= q\frac{{\mathbf r}\times {\mathbf n}}
{r(r - {\mathbf n} \cdot{\mathbf r})} \label{d_str}
\end{equation}
where the unit vector $\mathbf n$ determines the direction of a
string $S_{\mathbf n}$ passing from the origin of coordinates to
$\infty$ \cite{Dir}, and the Schwinger's choice is
\begin{equation}
{\mathbf A^{SW}}= \frac{1}{2}\bigl({\mathbf A}_{\mathbf n}+
{\mathbf A}_{-\mathbf n} \bigr), \label{sw}
\end{equation}
with the string being propagated from $-\infty$ to $\infty$ \cite{Sw_1}. Both
vector potentials yield the same magnetic monopole field, however the
quantization is different. The Dirac condition is $2\mu=p$, while the Schwinger
one is $\mu=p, \; p\in \mathbb Z$.

These two strings are members of a family $\{{\mathbf S}^{\kappa}_{\mathbf n} \}$ of  {\it weighted
strings}, which magnetic field is given by
\begin{align}
&{\mathbf h}^{\kappa}_{\mathbf n}=  \kappa{\mathbf h}_{\mathbf n}
+ (1-\kappa){\mathbf h}_{-\mathbf n} \label{str}
\end{align}
where $\kappa$ is the weight of a semi-infinite Dirac string. The respective vector potential reads
\cite{NF}
\begin{align}
{\mathbf A}^\kappa_{\mathbf n} = \kappa {\mathbf A}_{\mathbf n} + (1 -
\kappa){\mathbf A}_{-\mathbf n},
\label{A_1}
\end{align}
and since ${\mathbf A}^\kappa_{-\mathbf n}={\mathbf A}^{1-\kappa}_{\mathbf n}$, we obtain the following
equivalence relation: ${S}^{\kappa}_{-\mathbf n} \simeq{S}^{1-\kappa}_{\mathbf n} $.

Two arbitrary strings $S^\kappa_{\mathbf n}$ and $S^\kappa_{\mathbf n'}$ are related by the gauge transformation
\begin{eqnarray}
A^{\kappa'}_{\mathbf n'}= A^\kappa_{\mathbf n}+ d\chi. \label{ag2a}
\end{eqnarray}
and vice versa. Then an arbitrary transformation of the strings  $S^\kappa_{\mathbf n} \rightarrow
S^{\kappa'}_{\mathbf n'}$ can be realized as combination $S^\kappa_{\mathbf n} \rightarrow
S^{\kappa}_{\mathbf n'}$ and $S^\kappa_{\mathbf n} \rightarrow S^{\kappa'}_{\mathbf n}$, where the first transformation preserving the weight of the string is rotation, and the second one results in changing of the weight string $\kappa \rightarrow \kappa'$ without changing its orientation defined by $\mathbf n$.

Let denote by $\mathbf n'= g\mathbf n , g\in\rm SO(3)$, the left action of the
rotation group induced by $S^\kappa_{\mathbf n} \rightarrow S^\kappa_{\mathbf
n'}$. From rotational symmetry of the theory it follows this gauge
transformation can be undone by rotation $\mathbf r \rightarrow  \mathbf r g$
as follows \cite{Wu2,Jac,NF}:
\begin{align}
&A^{\kappa}_{\mathbf n'}(\mathbf r)= A^{\kappa}_{\mathbf n}(\mathbf r')= A^\kappa_{\mathbf n}(\mathbf r)+
d\alpha(({\mathbf r}; g)) , \label{g_0}\\
&\alpha(\mathbf r;g)= e \int_{\mathbf
r}^{\mathbf r'} \mathbf A^\kappa_{\mathbf n}(\boldsymbol \xi)
\cdot d \boldsymbol \xi, \quad \mathbf r' =  \mathbf r g
\label{g_1c}
\end{align}
where the integration is performed along the geodesic
$\widehat{\mathbf  r \,\mathbf r'}\subset S^2$.

Now returning to the transformation $S^\kappa_{\mathbf n} \rightarrow S^{\kappa'}_{\mathbf n}$ we obtain
\begin{eqnarray}
&&A^{\kappa'}_{\mathbf n} = A^{\kappa}_{\mathbf n} -
d\chi_{\mathbf n}, \label{A_02a}\\
 &&d\chi_{\mathbf n} = 2q{(\kappa' - \kappa)}\frac{(\mathbf r \times
\mathbf n)\cdot d\mathbf r}{r^2- (\mathbf n \cdot \mathbf r)^2},
\label{A_02}
\end{eqnarray}
$\chi_{\mathbf n}$ being polar angle in the plane orthogonal to ${\mathbf n}$. It is easy to see that this type of transformations can be undone by  combination of the inversion $\mathbf r\rightarrow -\mathbf r$ and  $\mu\rightarrow -\mu$. In particular, if $\kappa' = 1-\kappa$ we obtain the mirror string: $S^\kappa_{\mathbf n} \rightarrow S^{\kappa}_{-\mathbf n}\simeq S^{1-\kappa}_{\mathbf n} $.

Taking into account the spherical symmetry of the system, the vector potential can be considered as living on the two-dimensional sphere of the given radius $r$ and being taken as \cite{Wu1,Wu2}
\begin{eqnarray}
{\mathbf A_N} =  q\frac{1-\cos{\theta}}{r\sin{\theta}}\;
\hat{\mathbf e}_{\varphi}, \quad {\mathbf A_S} =
-q\frac{1+\cos\theta}{r\sin\theta}\; \hat{\mathbf e}_{\varphi}
\label{eq1b}
\end{eqnarray}
where $(r,\theta,\varphi)$ are the spherical coordinates, and while
${\mathbf A_{N}}$ has singularity on the south pole of the
sphere, ${\mathbf A_{S}}$ on the north one. In the overlap of
the neighborhoods  covering the sphere $S^2$ the potentials
${\mathbf A_N}$ and ${\mathbf A_S}$  are related by the following gauge
transformation:
\[
A_S = A_N - 2qd\varphi.
\]
This is the particular case of (\ref{A_02a}), (\ref{A_02a}) when $\kappa = 0$ and $\kappa' =1$.

Choosing the vector potential as $\mathbf A_N$ we have
\begin{eqnarray}
&&J_{\pm}= e^{\pm i\varphi}\bigg(\pm\frac{\partial}{\partial\theta}
+i\cot\theta \frac{\partial}{\partial\varphi} -
\frac{\mu\sin\theta}{1+\cos\theta} \bigg),\\
&&J_3=-i\frac{\partial}{\partial\varphi} - \mu,\\
&&{\mathbf J^2} =-\frac{1}{\sin{\theta}}
\frac{\partial~}{\partial\theta}\left(\sin{\theta}
\frac{\partial~}{\partial\theta}\right) -
\frac{1}{\sin^2{\theta}}\frac{\partial^2~}{\partial\varphi^2} + \nonumber\\
&&+i\frac{2\mu}{1 +\cos{\theta}}\frac{\partial~}{\partial\varphi}
+\mu^2\frac{1 - \cos{\theta}}{1 + \cos{\theta}} +\mu^2 \label{eq7}
\end{eqnarray}
where  $J_{\pm} = J_x \pm iJ_y$ are the raising and the lowering operators
for $J_3 = J_z$.

Writing Schr\"odinger's equation
\begin{equation}
\hat H\Psi = E \Psi, \label{eq01}
\end{equation}
in the spherical coordinates, and putting $\Psi= R(r)Y(\theta,\varphi)$
into Eq. (\ref{eq01}), we get for the angular part the following equation:
\begin{eqnarray}
&&{\mathbf J^2}Y(\theta,\varphi) = \ell({\ell} + 1)Y(\theta,
\varphi). \label{eq7a}
\end{eqnarray}

Assuming
\begin{eqnarray*}
Y =e^{i(m+\mu)\varphi}z^{p}(1-z)^{q}F(z),
\end{eqnarray*}
 where $z=(1-\cos\theta)/2$ and $m$ is an eigenvalue of $J_3$, we obtain
the resultant equation in the standard form of the hypergeometric
equation,
\begin{eqnarray}
z(1-z)\frac{d^2F}{dz^2} +\bigl(c-(a+b+1)z\bigr)\frac{dF}{dz}-abF=0
\label{hyp1}
\end{eqnarray}
where
\begin{align}
&a = p + q - \ell, \; b = p + q + \ell + 1, \; c = 2p + 1, \label{sols}\\
&(p + q)(p - q) = m\mu \label{sols_1}
\end{align}

As it is known the hypergeometric function $F(a,b;c;z)$ reduces to a polynomial of
degree $n$ in $z$ when $a$ or $b$ is equal to $-n, \;(n = 0,1,2, \dots)$, and
the respective  solution of Eq.(\ref{hyp1}) is of the form \cite{abr,and}
\begin{eqnarray}
F=z^{\rho}{(1-z)}^{\sigma} p_n(z) \label{pol}
\end{eqnarray}
where $p_n(z)$ is a polynomial in $z$ of degree $n$. Here we are looking for the solutions, like this of
the Schr\"odinger equation (\ref{eq7a}). The requirement of the wave function
being single valued force us to take $\alpha = m +\mu$ as an integer and
general solution is given by
\begin{equation}\label{gsol_1}
 Y_{\ell}^{(\mu,n)}=e^{i\alpha\varphi}Y_n^{(\delta,\gamma)}(u),
\end{equation}
where $u = \cos\theta$, and
\begin{equation}\label{pol_2}
Y_n^{(\delta,\gamma)}(u) =C_n\,(1-u)^{\delta/2}(1+u)^{\gamma/2}
P_n^{(\delta,\gamma)}(u),
\end{equation}
$P_n^{(\delta,\gamma)}(u)$ being the Jacobi polynomials, and the normalization constant $C$ is given by
\begin{equation*}
C_n=\Bigg(\bigg|\frac{2 \pi \, 2^{\delta +\gamma +1}\Gamma(n +\delta +1)
\Gamma(n +\gamma +1)}{\Gamma(n+1)\Gamma(n +\delta +\gamma+1)}\bigg|    \Bigg)^{-1/2}
\end{equation*}

It follows from Eqs. (\ref{sols}), (\ref{sols_1}) four different cases (we set $\beta = m-\mu$):
\begin{align*}
&\stackrel{{\quad}+\;(\mu,n)}{Y_{\ell\pm \mu}}
=e^{i\alpha\varphi}Y_n^{(\alpha,\beta)}\left\{\begin{array}{l}
m= \ell -n,\;\ell+\mu \in {\mathbb Z}_{+}\\
m= -\ell- n-1, \; \ell-\mu \in {\mathbb Z}_{+}
  \end{array}\right .\\
&\stackrel{{\quad}-\;(\mu,n)}{Y_{\ell\pm \mu}}
=e^{i\alpha\varphi}Y_n^{(-\alpha,-\beta)}\left\{\begin{array}{l}
m= \ell+ n+1, \; \ell+\mu \in {\mathbb Z}_{+}\\
m= n-\ell,\;\ell-\mu \in {\mathbb Z}_{+}
  \end{array}\right .
\end{align*}

The functions $\{\stackrel{{\quad}+\;(\mu,n)}{Y_{\ell\pm \mu}}\}$
form the basis in the indefinite metric Hilbert space for the nonunitary
representation $D^-(\ell \pm\mu,\mu)$ bounded above, and the set of functions
$\{\stackrel{{\quad}-\;(\mu,n)}{Y_{\ell\pm \mu}}\}$  for the
representation $D^+(\ell \pm\mu,\mu)$ bounded below. Notice that for $\mu$ being an arbitrary
valued the representations corresponding to $\ell+\mu \in {\mathbb Z}_{+}$
and $\ell-\mu \in {\mathbb Z}_{+}$ are not equivalent.

The wave functions
\begin{equation}\label{eq9a}
  Y_{\ell}^{(\mu,n)}=\Big\{\stackrel{{\quad}+\;(\mu,n)}{Y_{\ell\pm \mu}}
,\stackrel{{\quad}-\;(\mu,n)}{Y_{\ell\pm \mu}}\Big\}
\end{equation}
where the upper/lower sign corresponds to the choice of
$(\ell+\mu)$ or $(\ell-\mu) $, form a complete set of
orthonormal solutions with indefinite metric
\begin{align*}
&\eta_{np}= \left\{\begin{array}{l}
\delta_{np},\; {\rm if}\; 2\ell -n >0\\
\delta_{np}(-1)^{n+1}\,{\rm sgn}(\sin2\pi\ell),\; {\rm if}\; n -2\ell >0,
  \end{array}\right.
\end{align*}
one can set $m = p \pm \ell \; (p= 0, \pm 1,\pm 2,\dots)$, where the upper sign corresponds to the
representation defined by $\ell + \mu$  and the lower one to $\ell -\mu$.

Similar consideration can be done for the vector potential
$\mathbf A_S$. In this case $\beta =m-\mu\in \mathbb Z$ and the
corresponding wave functions
\begin{align*}
&\stackrel{{\quad}+\;(\mu,n)}{Y_{\ell\pm \mu}}
=e^{i\beta\varphi}Y_n^{(\alpha,\beta)}\left\{\begin{array}{l}
m= -\ell- n-1, \; \ell+\mu \in {\mathbb Z}_{+}\\
m= \ell -n,\;\ell-\mu \in {\mathbb Z}_{+}
  \end{array}\right .\\
&\stackrel{{\quad}-\;(\mu,n)}{Y_{\ell\pm \mu}}
=e^{i\beta\varphi}Y_n^{(-\alpha,-\beta)}\left\{\begin{array}{l}
m= n-\ell,\;\ell+\mu \in {\mathbb Z}_{+} \\
m= \ell+ n+1, \; \ell-\mu \in {\mathbb Z}_{+}
  \end{array}\right .
\end{align*}
form a complete set of orthonormal basis for the nonunitary
representation $D^{\pm}(\ell \pm\mu,-\mu)$.

Thus we find the following series of the representations:
\begin{align*}
&\ell+\mu \in {\mathbb Z}_{+} \Rightarrow\left\{\begin{array}{l}
D^-(\ell +\mu,\mu): \; m= \ell -n\\
D^+(\ell +\mu,\mu): \; m= n + \ell +1\\
D^-(\ell +\mu,-\mu): \;m= -\ell- n-1\\
D^+(\ell +\mu,-\mu): \;m= n-\ell\\
 \end{array}\right .\\
&\ell-\mu \in {\mathbb Z}_{+} \Rightarrow\left\{\begin{array}{l}
D^-(\ell -\mu,\mu): \; m= -\ell -n -1\\
D^+(\ell -\mu,\mu): \; m= n - \ell\\
D^-(\ell -\mu,-\mu): \;m=\ell - n \\
D^+(\ell -\mu,-\mu): \;m= n +\ell +1\\
 \end{array}\right .
\end{align*}
where $n = 0,1,2,\dots$. Taking into account the following restriction:  $\ell(\ell+1)-\mu^2 \geq 0$,
emerging from the Schr\"odinger equation, the allowed values of $l$ are found to be
\begin{align*}
&\ell+ \mu \in {\mathbb Z}_{+}\,\Rightarrow \, \ell= -\mu + [2\mu] +k, \quad k = 0,1,2,\dots \\
&\ell- \mu \in {\mathbb Z}_{+}\,\Rightarrow \, \ell= \mu + k, \quad k = 0,1,2,\dots
\end{align*}
where $[2\mu]$ denotes the integer part of $2\mu$.

The function $Y_{\ell}^{(\mu,n)}$ being a member of the family
$\{Y_{\kappa,\ell}^{(\mu,n)}\}$ of the so-called {\it weighted
monopole harmonics} such that \cite{NF}
\begin{equation}
Y_{\kappa,\ell}^{(\mu,n)}= {e}^{-i2\kappa\mu\varphi}
Y_\ell^{(\mu,n)}, \quad  2\kappa \mu \in \mathbb Z\label{eq2_a}
\end{equation}
is a solution of the Schr\"odinger equation corresponding to the choice of the
vector potential as
\[
{\mathbf A}^{\kappa} =\kappa{\mathbf A}_{S} + (1-\kappa){\mathbf
A}_{N},
\]
For a given $\mu$ a weight $\kappa$ is quantized parameter in units of $\mu$, and
in particular cases $\kappa =1$ and $\kappa=1/2$ it yields the Dirac and Schwinger selectional rules
respectively.

Since the set of weighted monopole harmonics $\{Y_{\kappa,\ell}^{(\mu,n)} \}$ forms the orthonormal basis
in the indefinite metric Hilbert space of the irreducible infinite-dimensional nonunitary representation
$D^+(\ell,\mu)\otimes D^-(\ell,\mu)$, any solution of the Schr\"odinger's equation (\ref{eq7a}) can be
expanded as
\begin{equation}
\Psi = \sum_{ln} C_{ln}Y_{\kappa,\ell}^{(\mu,n) } \label{eq3d}
\end{equation}
where $\mu$ is an {\em arbitrary parameter}.

When $n+\alpha$, $n+\beta$ and $n+\alpha+\beta$ all are integers $\geq 0$ and $\kappa =0$ the weighted
monopole harmonics are reduced to the {\em monopole harmonics} introduced by  Wu and Yang \cite{Wu2}. The
imposed here restrictions on the values of $n,\alpha$ and $\beta$ yield the finite-dimensional unitary
representation of the rotation group and Dirac quantization condition.

\section{Discussion and concluding remarks}

Involving infinite-dimensional representations of the rotation group, we have deduced a consistent
pointlike monopole theory  with an arbitrary magnetic charge. It follows from our approach  a generalized
quantization condition $2\kappa\mu = 0, \pm 1,\pm 2,\dots $, that can be considered as quantization of the
weight string $\kappa$ instead of the monopole charge. In particular cases $\kappa =1$ and $\kappa=1/2$ we
obtain the Dirac and Schwinger selectional rules respectively.

Using infinite-dimensional representations of the rotation group one has to
employ the indefinite metric Hilbert space. What makes difference between our
approach and others recently have been developed in the growing number of
papers on the subject of $PT$-symmetric quantum mechanics, is absence of
``negative probability''. Thus we avoid the problem of the negative
probability and preserve the standard probabilistic interpretation of the
quantum mechanics.

The other important aspect of the Dirac monopole problem is the gauge-invariant
algebra of translations. As is known, the Jacobi identity fails and for the
finite translations one has \cite{Jac,Gr}
\begin{align}
\bigl(U_{\mathbf a}U_{\mathbf b}\bigr)U_{\mathbf c}
=\exp(i\alpha_3(\mathbf r ;\mathbf a, \mathbf b,\mathbf c)) U_{\mathbf
a}\bigl(U_{\mathbf b}U_{\mathbf c}\bigr) \label{as}
\end{align}
where $\alpha_3= 4\pi \mu\,\mod 2\pi \mathbb Z$, if the monopole is enclosed by
the simplex with vertices $(\mathbf r,\mathbf r +\mathbf a, \mathbf r+ \mathbf
a +\mathbf b,\mathbf r+ \mathbf a+ \mathbf b +\mathbf c)$ and zero otherwise
\cite{Jac}. For the Dirac quantization condition being satisfied $\alpha_3 = 0
\mod 2\pi \mathbb Z$, and (\ref{as}) provides an associative representation of
the translations, in spite of the fact that the Jacobi identity continues to
fail. Since a conventional quantum mechanics deals with linear Hilbert space
and hence with associative algebra of observables, avoiding of Dirac's the
quantization condition forces us to go beyond the standard quantum mechanical
approach and introduce {\it nonassociative algebra of observables}
\cite{Jac,Gr,Gr1,G1,G2}. This work is in a progress.


\begin{thebibliography}{99}

\bibitem {Dir} P. A. M. Dirac, {\em Quantized Singularities in the Electromagnetic Field,   Proc. Roy. Soc.
Lond.} {\bf A 133} (1931) 60 .

\bibitem {Sw_1} J. Schwinger, {\em Magnetic Charge and Quantum Field Theory,}  \pr{144} {1966} {1087}.

\bibitem {Wu1} T. T. Wu and C. N. Yang, {\em Concept of nonintegrable phase factors and global formulation
of gauge fields,} \prd {12} {1975} {3845}.

\bibitem {Wu2} T. T. Wu and C. N. Yang, {\em Dirac Monoploe without Strings: Monopole Harmonics,} \npb {
107} {1976} {365}.

\bibitem {Jac} R. Jackiw, {\em , Three-Cocycle in Mathematics and Physics} \prl{54} {1985} {159}.

\bibitem {Gr} B. Grossman, {\em A 3-Cocycle in Quantum Mechanics,} \plb{152}{1985} {93}.

\bibitem {Gr1} B. Grossman, {\em Three-cocycle in quantum mechanics.II,} \prd{33}{1986} {2922}.

\bibitem {G1} Y.-S. Wu and A.Zee, {\em Cocyle and Magnetic Monopole,} \plb {152} {985} {98} .

\bibitem {G2} D.G. Boulware, S. Deser and B. Zumino, {\em Absence of 3-cocycles in the Dirac Monopole
Problem,} \plb {153} {1985} {307}.

\bibitem {G3} M. Nakahara, {\it Geometry, Topology and Physics} (IOP,
London, 1990).

\bibitem {Ch} T.-P. Cheng and L.-F. Li, {\it Gauge theory of elementary particle} (Clarendon, Oxford,
1984).

\bibitem {Gol1} A.S. Goldhaber, {\em Role of Spin in the Monopole Problem,} \prb{140} {1965)} {1407}.

\bibitem {Gol2} A.S. Goldhaber, {\em Connection of Spin and Statistics for Charge-Monopole Composites,}
\prl {36} {1976}{1122}.

\bibitem {Zw_1} D. Zwanziger, {\em Quantum Field Theory of Particles with Both Electric and Magnetic
Charges,}  \prd{176} {1968} {1489}.

\bibitem {Zw_2} D. Zwanziger, {\em Local-Lagrangian Quantum Field Theory of Electric and Magnetic Charges,}
\prd{3} {1971}{880}.

\bibitem {H} A. Hurst, {\em Charge Quantization and Nonintegrable Lie Algebras,} \ap {50} {1968} {51}.

\bibitem  {NF} A.I. Nesterov and F. Aceves de la Cruz, {\em Magnetic monopoles with generalized
quantization condition,} \pla{302} {2002} {253} [\hepth{0208210}].

\bibitem {N1} A.I. Nesterov, {\it Principal $Q$-bundles.} In: {\it Non
Associative Algebras and Its Applications}, ed. R. Costa,H. Cuzzo,
Jr. A. Grishkov and L.A.  Peresi (Marcel Dekker, New York, 2000).

\bibitem  {N2} A.I. Nesterov, {\em Principal Loop Bundles: Toward Nonassociative Gauge Theories,} \ijtp
{40} {2001} {339}.

\bibitem {BB} C. M. Bender and S. Boettcher, {\em Real Spectra in Non-Hermitian Hamiltonians Having PT
Symmetry,} \prl{80}{1998}{5243} [physics/{9712001}].

\bibitem {BBM} C.Bender, S. Boettcher and P. Meisinger, {\em PT-Symmetric Quantum Mechanics,} \jmp
{40}{1999} {2201} [\quantph{9809072}].

\bibitem {BBJ} C. M. Bender, D.C. Brody and H.F. Jones, {\em Complex Extension of Quantum Mechanics,}
\prl{89}{2002}{270401} [\quantph{0208076}].

\bibitem {M1}  A. Mostafazadeh, {\em Pseudo-Hermiticity for a Class of Nondiagonalizable Hamiltonians,}
\jmp{43}{2002} {6343}; Erratum-ibid. {\bf 44} (2003) 943 [\Math{ph}{0207009}] .

\bibitem {M2}  A. Mostafazadeh, {\em Pseudo-Hermiticity versus PT Symmetry: The necessary condition for the
reality of the spectrum of a non-Hermitian Hamiltonian,} \jmp {43}{2002} {205} [\Math{ph}{0107001}].

\bibitem {M3}  A. Mostafazadeh, {\em Pseudo-Hermiticity versus PT-Symmetry II: A complete characterizatio n
of non-Hermitian Hamiltonians with a real spectrum,} \jmp {43}{2002} {2814} [\Math{ph}{0110016}].

\bibitem {M4} A. Mostafazadeh, {\em Pseudo-Unitary Operators and Pseudo-Unitary Quantum
Dynamics} [\Math{ph}{0302050}].

\bibitem {M5} A. Mostafazadeh, {\em Generalized PT -, C- and CPT -Symmetries, Position
Operators, and Localized States of Klein-Gordon Fields}, [\quantph{0307059}].

\bibitem {Sol} L. Solombrino, {\em Weak pseudo-Hermiticity and antilinear commutant,} \jmp{43}{2002} {5439}
[\quantph{0203101}].

\bibitem {BSS} A. Blasi, G. Scolaric and L. Solombrino, {\em Pseudo-Hermitian Hamiltonians, indefinite
inner product spaces and their symmetries,} [\quantph{0310106}].


\bibitem {RM} A. Ram\'irez and B. Mielnik,{\em The Challenge of non-hermitian structures in physics, Rev.
Mex. Phys.} {\bf 49 S2} (2003) 130 [\quantph{0211048}].


\bibitem {Pauli} W. Pauli, {\em On Dirac's New Method of Field Quantization,} \rmp {15} {1943} {175}.

\bibitem {GSH} I. M. Gel'fand and G. E. Shilov, {\em Generalized Functions,} Vol. I (Academic Press, New
York, 1964).

\bibitem{AG} M. Andrews and J. Gunson, {\em Complex Angular Momenta and Many-Particle States. I. Properties
of Local Representsations of the Rotation Group}, \jmp{5}{1964}{1391}.

\bibitem {BL} E.G. Beltrami and G. Luzatto, {\em Rotation Matrices Corresponding to Complex Angular
Momenta,} \nc{XXIX}{1963}{1003}.

\bibitem {S} S.S. Sannikov, {\em Representations of the Rotation Group with Complex Spin,}
\sjnp{2}{1966}{407}.

\bibitem {S1} S.S. Sannikov, {\em Infinite-domensional Representations of the Rotation Group,}
\sjnp{6}{1968}{788}.


\bibitem {S2} S.S. Sannikov, {\em New Representations of the Lie Algebra of the Rotation Group,}
\sjnp{6}{1968}{939}.


\bibitem {W}  B. G. Wybourne, {Classical groups for Physicists}
(Wiley, New York, 1974).


\bibitem {and} G. E. Andrews, R. Askey, R. Roy, {\em Special
functions} (Cambridge, Cambridge University Press, 2000).

\bibitem {abr}  {\em Handbook of Mathematical Functions}, ed. M. Abramowitz
and I. A. Stegun (Dover, New York, 1965).


\end{thebibliography}
\end{document}